\documentclass[twocolumn,showpacs,prl,10pt,superscriptaddress]{revtex4}
\usepackage[english]{babel}
\usepackage{amsmath,amssymb}
\usepackage{times}
\usepackage{epsfig}

\def\tcw{90mm}

\def\gsim{\mathrel{\scriptstyle{\buildrel > \over \sim}}}
\def\lsim{\mathrel{\scriptstyle{\buildrel < \over \sim}}}

\begin{document}
\title{Thermodynamics of the planar Hubbard model}
\affiliation{J.\ Stefan Institute, SI-1000 Ljubljana, Slovenia}
\author{J. \surname{Bon\v ca}}
\author{P. \surname{Prelov\v sek}}
\affiliation{J.\ Stefan Institute, SI-1000 Ljubljana, Slovenia}
\affiliation{Faculty of Mathematics and Physics, University of Ljubljana,
SI-1000 Ljubljana, Slovenia}
%\preprint{IJS-TP 02/xx}
\date{\today}

\begin{abstract}
The thermodynamic properties: specific heat, entropy, spin
susceptibility $\chi_s$ and charge susceptibility $\chi_c$ are studied
as a function of temperature and doping within the two-dimensional
Hubbard model with various $U/t=4 - 12$.  Quantities are calculated
using the finite-temperature Lanczos method with additional
phase-averaging for a system of $4\times 4$ sites. Results show that
the entropy at low $T$ reaches a maximum near half-filling at the
electron density $n \sim 1\pm 0.15$ in the whole regime of studied
$U/t$. The pseudogap in $\chi_s(T)$ becomes clearly pronounced for
$U/t \geq 8$ while $\chi_c$ shows a maximum close to half-filling.
The relation of results to those within the $t$-$J$ model and to
experiments is discussed.

\end{abstract}
\maketitle

The Hubbard model is the simplest prototype Hamiltonian for correlated
electrons. It has been and still remains the subject of numerous
theoretical investigations in connection with the metal-insulator
transition \cite{imad}, the interplay between the magnetism and
the itinerant character of electrons, and possible superconductivity
emerging solely from the electronic mechanism. A particular
attention has been devoted to the two-dimensional model (2D) on a
square lattice, expected to capture the physics of superconducting
cuprates. A lot of effort has been put into the numerical studies of the
ground state properties, using various quantum Monte Carlo (QMC)
methods \cite{dago}.

On the other hand, there are rather few studies of the 2D Hubbard
model at finite $T>0$, in particular away but close to the half
filling, i.e. at the electron densities $n \sim 1$. In the latter
regime the minus-sign problem prevents the application of the QMC
method at low $T$ in large systems \cite{dago}. Gross features of the
specific heat $C_V(T)$ have been obtained via the internal energy
$E(T)$ using the QMC \cite{duffy}. Results reveal the evidence of at
least two energy scales at large $U/t \gg 1$, the larger one
representing the upper Hubbard band. The behavior at low $T$ shows a
marked difference between an insulator at half filling $n=1$ with
$C_V(T)\propto T^2$, and an anomalous metal at finite hole doping
$n_h=1-n > 0$ (or analogous electron doping). Within the metallic
regime the QMC method was so far not able to reach temperatures below
the exchange scale $J \sim 4t^2/U$, which sets up an characteristic
energy of spin dynamics and is thus essential for establishing the
low-$T$ physics at low doping. The uniform spin susceptibility
$\chi_s(T)$ has also been calculated \cite{moreo} using QMC, with even
larger restrictions (smaller systems) at finite doping $n \neq 1$, and
by Dynamical Cluster Approximation \cite{jarrell}. On
the other hand, low-$T$ properties of the Hubbard model with $U \gg t$
are believed to map well on the properties of the $t$-$J$ model which
is projected on the basis space without doubly occupied sites.
Several static and dynamic properties of the planar $t$-$J$ model have
been recently calculated and followed well into the regime $T<J$ using
the finite temperature Lanczos method (FTLM) \cite{jplan,jprev}. Two
most relevant conclusions on the thermodynamic properties of the 2D
$t$-$J$ model \cite{jpterm} are: a) normal-state entropy density
$s(T<J)$ is maximum at the 'optimum' hole doping $n_h \sim n_h^*$
where $n_h^* \sim 0.15$ at $J/t=0.3$, b) a pseudogap temperature
$T^*(n_h)$, experimentally (among alternatives) defined with the
maximum in the uniform spin susceptibility $\chi(T)$
\cite{torr,batl,imad}, shows up also in the $t$-$J$ model where
$T^*(n_h)$ decreases with doping and vanishes at the 'optimum' one, c)
even at quite low $T \ll J$ and in the 'underdoped' regime $n_h<n_h^*$
some thermodynamic properties are close to the behavior of a
semiconductor-like nondegenerate fermion gas \cite{prel}.

Our aim is to obtain thermodynamic results within the planar Hubbard
model, which is numerically (for an exact diagonalization approach)
clearly more demanding relative to the $t$-$J$ model. We list some
relevant questions which we address in the following: a) are there any
qualitative differences between the thermodynamic properties of the
planar $t$-$J$ model and the Hubbard model at large $U/t$, b) how does
the entropy 'optimum' doping shift with decreasing $U/t$, c) is there
a pseudogap scale also at smaller $U/t$.

We investigate the Hubbard model given by
\begin{equation}
H=-t\sum_{\langle ij\rangle s}( c_{is}^\dagger
c_{js}^{\phantom{\dagger}}+\text{H.c.})+ U\sum_{i}
n_{i\uparrow}n_{i\downarrow}, \label{eq1}
\end{equation}
where $c^\dagger_{is}(c_{is})$ and $n_{is}$ are creation
(annihilation) and number operators for electrons, respectively, and
the sum $\langle ij\rangle$ runs over pairs of nearest-neighbor sites.
We limit our calculations to $U/t =4,8,12$, where values range from
the modest $U<W$, smaller than the bandwidth $W=8t$, to the strong
correlation regime $U>W$. Note that the latter case corresponds to the
physics of cuprates where the spin exchange $J \sim 4 t^2/U \sim
0.3~t$.

We study numerically the Hubbard model on a square lattice using the
FTLM \cite{jplan,jprev}, based on the Lanczos procedure of exact
diagonalization and a random sampling over initial
wavefunctions. The advantage in the case of thermodynamic quantities
is that they can be expressed solely in terms of a grand-canonical
average of conserved quantities ($k_B=1$), i.e.,
\begin{equation}
\langle f\rangle = {\rm Tr} f(N_e,S_z,H) e^{-(H-\mu N_e)/T}
/{\rm Tr}e^{-(H-\mu N_e)/T}, \label{eq2}
\end{equation}
where $N_e$, $S_z$ and $\mu$ refer to the number of electrons, the
total spin and the chemical potential, respectively. In the case of
quantities as in Eq.(\ref{eq2}), the FTLM does not require the storage
of Lanczos eigenfunctions, but only of Lanczos eigenenergies
$\epsilon_j^n$, where $j=0,\cdots, M$ ($M$ represents the number of
Lanczos steps) while $n=1,\cdots,R$ runs over random initial
Lanczos wavefunctions. We refer for the details of the method to
Refs.\cite{jpterm,jprev}. Using FTLM in the above way we are able to
investigate the model on the lattice of $N=4 \times 4=16$ sites with
periodic boundary conditions. 

The main limitation to the validity of results comes from finite-size
effects. The latter can be substantially reduced by employing the
boundary condition (flux) averaging \cite{poil}.  In a system with
periodic boundary conditions the latter is achieved by introducing the
uniform vector potential $\vec \theta$ modifying the hopping elements
$t \to \tilde t_{ij} = t ~\mathrm {exp}(i \vec \theta \cdot \vec
r_{ij})$.  We use furtheron $N_t$ uniformly spaced phases $\vec
\theta$ instead of a fixed $\theta=0$. In this way results are
essentially improved at lower $U<W$. This is particularly evident for
noninteracting electrons with $U=0$, where results on small lattices
otherwise reveal pronounced finite-size effects. In this case, using
$N_t \gg 1$ most properties discussed here become exact even on a
finite-size lattice.

Still the main restriction in the thermodynamic validity of our
results comes from finite-size effects which show up at $T<T_{fs}$
where they start to dominate results \cite{jprev}. In the particular
parameter space $U/t=4-12$, the 'optimum' cases are at $n \sim 1\pm
n_h^*$ with $n_h^* \sim 0.15$ (coinciding with largest entropy
$s=s_{max}$) where $T_{fs}/t \sim 0.1-0.15$. On the other hand,
$T_{fs}$ increases towards $n=1$ and $n \to 0,2$, respectively
\cite{jprev}. Since the properties of the Hubbard model (\ref{eq1}) on
a bipartite lattice are symmetric around half-filling we present
results only for the hole-doped regime $n_h=1-n \geq 0 $.

Using Eq.(\ref{eq2}) we directly evaluate within FTLM the electron
density $n=\langle N_e \rangle/N$, the entropy density $s$, expressed
as
\begin{equation}
s=\ln\Omega/N+(\langle H\rangle-\mu\langle N_e\rangle)/NT,
\label{eq3}
\end{equation}
and the spin susceptibility $\chi_s=\langle (S_z)^2\rangle/N
T$. Quantities calculated as functions of $\mu$ and $T$ can be
consequently presented as well as in terms of $n$ and $T$. Using above
quantities we also evaluate the specific heat $C_V=T(\partial
s/\partial T)_{\mu}$ and the charge susceptibility - electron
compressibility $\chi_c=(\partial n/\partial \mu)_{T}$.

Let us first discuss FTLM results for an overall behavior of the
specific heat $C_V(T)$ (per unit cell), as shown in Fig.~1 for $U/t=0
- 12$ in the whole relevant $T$ regime. At high $T>0.5~t$ our FTLM
results in general agree with those obtained previously with the QMC
method \cite{duffy}. The advantage of FTLM is that we can reach lower
$T\sim T_{fs} \sim 0.1~t$, well below the exchange scale $T \ll J \sim
4t^2/U$. The main message of Fig.~1 is that $C_V$ reveals the
existence of (at least) two energy scales which are well separated for
$U\gg t$, i.e. for $U =12~t$. The upper maximum is related to
excitations within the upper Hubbard band and is well pronounced near
half-filling. For a larger doping, i.e. for $n<0.85$, these
excitations merge with the lower Hubbard band.  At lower $U=4~t$, the
upper maximum is only weakly present even at $n=1$, and disappears at
smallest available doping $n_h=0.95$.  Note also that at $U=4~t$,
apart from $n=1$, $C_V$ merges even quantitatively with the
noninteracting result, $U=0$ (properties at $U=0$ in Figs.~1 - 4 are
calculated for an infinite lattice). When discussing the relation of
presented results to those within the $t$-$J$ model we point out that
the upper scale (upper Hubbard band) is projected out in the latter so
results for $C_V$ are typically different for $T>t$.  \cite{jpterm}.

\begin{figure}[htb]
\centering
\center{\epsfig{file=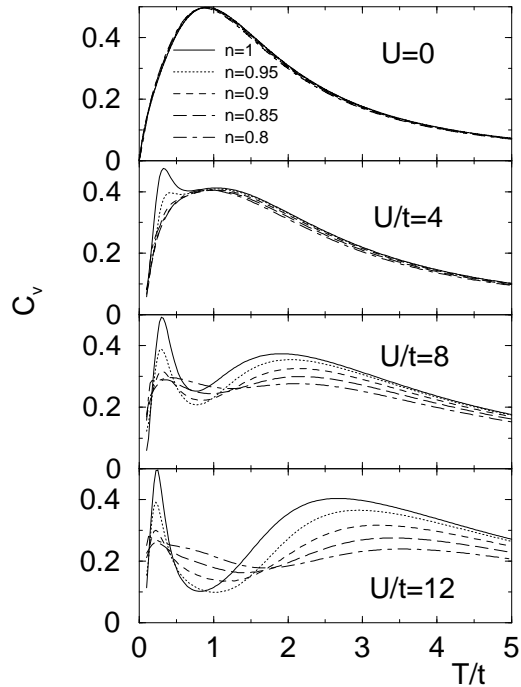,width=\tcw,angle=-90}}\\
\caption{Specific heat $C_V$ (per unit cell) vs. $T$ for various
electron densities $n$ near half-filling and different $U/t$.  $U=0$
result is calculated for an infinite lattice.}
\label{fig1}
\end{figure}

In the following we focus on the lower energy scale which is essential
for the understanding of quasiparticle and low-$T$ properties. In
Fig.~2 we show entropy density $s$ as a function of electron density
$n$ for different $U/t=0 - 12$ as well as for few lowest $T/t=0.1 -
0.3$.  First observation is that $U>0$ leads to an increase of $s$,
which is largest at an intermediate doping $n_h = n_h^* \sim 0.15$. As
expected, results for $U=12~t$ are even quantitatively close to the
ones within the $t$-$J$ model \cite{jpterm,jprev} with the
corresponding $J=0.3~t$ where the maximum $s$ has as well been
observed at $n_h^*\sim 0.15$ and such a doping has been identified as
an 'optimum' one. We should note that such a characterization of
'optimality' does not seem to be in conflict with the usual one
related to highest $T_c$ since experimentally in several cuprates the
maximum in $T_c$ and in the entropy \cite{coop} appear to be quite
close in doping. Plausibly, $n_h^*$ can be related to the most
frustrated case where the kinetic energy of holes (preferring an
ferromagnetic ordering) and the spin exchange (favoring
antiferromagnetism) are competing and therefore one could expect
$n_h^* \propto J/t$. Moreover, it is evident from Fig.~2 that the
'optimal' doping $n_h^*\sim 0.15$ is quite insensitive to $U$ in a
broad range $U/t=4-12$.

\begin{figure}[htb]
\centering
\center{\epsfig{file=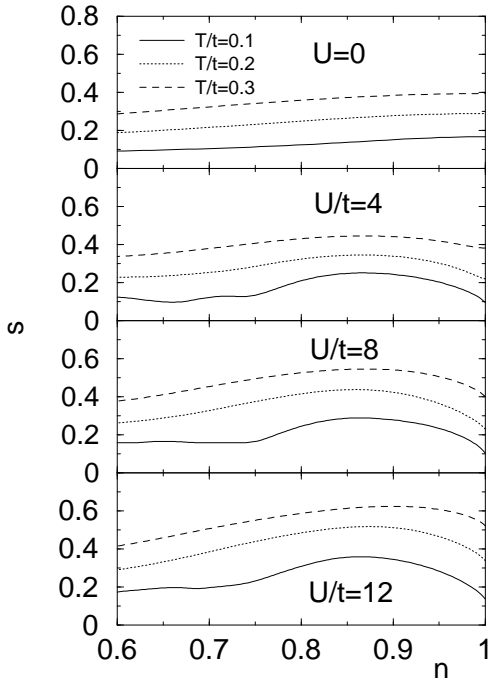,width=\tcw,angle=-90}}\\
\caption{Entropy density $s$ vs. electron density $n$ for low 
$T/t=0.1-0.3$ and different $U/t$.}
\label{fig2}
\end{figure}

In Fig.~3 we present results for the spin susceptibility $\chi_s(T)$
for various dopings close to half-filling $n=0.8-1.0$ and $U/t=0 -
12$.  We first note that here the phase averaging method brings
substantial improvement.  This is evident by comparing Fig.~3 with QMC
results on the same $4\times 4$ lattice \cite{moreo} obtained at a
fixed phase $\theta=0$.

As expected, the onset of $U>0$ leads to an increase of $\chi_s(T)$ at
lower $T<t$. It is however more interesting to follow the development
of pseudogap features with increasing $U/t$. One of experimental
definitions of the (large) pseudogap temperature is related to the
maximum $\chi_s(T=T^*)$ \cite{torr}. In fact, $T^*$ defined in this
way matches well with other experimentally established crossovers
\cite{batl,imad}. It has been found \cite{jpterm,jprev} that
$T^*(n_h)$ determined in this way within the $t$-$J$ model matches well
experiments. As foreseen from the mapping to the $t$-$J$ model with
$J=0.3~t$, we find in Fig.~3 essentially the same behavior for the
Hubbard model with $U/t=12$. On the other hand, the pseudogap maximum
becomes shallower for $U/t=8$, although the location $T^*(n)$ does not
seem to shift substantially. The pseudogap features disappear at
$U/t=4$.

\begin{figure}[htb]
\centering
\epsfig{file=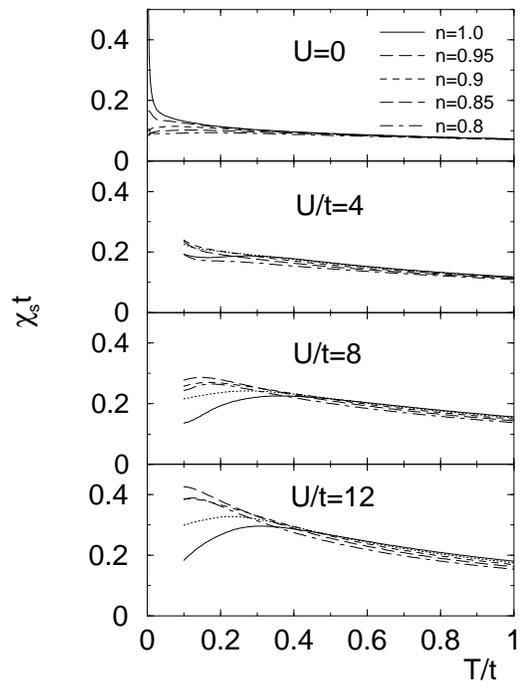,width=\tcw,angle=-90}
\caption{Spin susceptibility $\chi_s t$ vs. T for $n=0.8-1.0$ and different 
$U/t$.}
\label{fig3}
\end{figure}

Let us finally comment on results for the charge susceptibility
$\chi_c=dn/d\mu$, as presented in Fig.~4. For noninteracting electrons
at $U=0$, $\chi_c$ is essentially $T$-independent (except very close
to $n=1$, due to the van-Hove singularity) and is equal to the single
- electron density of states at the Fermi energy $\chi_c = {\cal
N}_F$. Well away from half-filling, i.e. in the 'overdoped' regime'
$n<0.8$, the effect of $U>0$ is only quantitative to reduce
$\chi_c$. This can be attributed to an overall decrease of the
effective density ${\cal N}(\epsilon)$ due to the transfer of states
into the upper Hubbard band. We also note in Fig.~4 that at the same
time $U>0$ leads to an even flatter variation of $\chi_c(n)$.

\begin{figure}[htb]
\centering
\epsfig{file=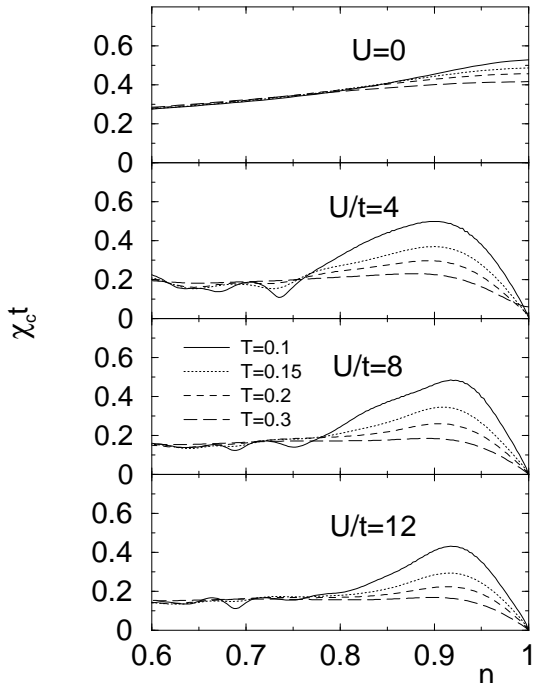,width=\tcw,angle=-90,clip=}
\caption{Charge susceptibility $\chi_c t$ vs. $n$ for 
$T/t=0.1-0.3$ and different $U/t$.}
\label{fig4}
\end{figure}

More indicative and challenging is the development within the
'underdoped' regime $n_h<n_h^*$, with a pronounced $T$ and doping
dependence. Very close to half-filling $n \sim 1$, we are at $T>
T_{fs}$ dealing with chemical potential $\mu$ within the charge
(Mott-Hubbard) gap.  A small density of charge carriers $n_h \ll 1$ in
this regime behaves as in a doped nondegenerate semiconductor, (as
established within the $t$-$J$ model at low doping \cite{prel}) where
\begin{equation}
n_h \sim P {\mathrm e}^{{\displaystyle -(\mu -\epsilon_v)/T}}.
\label{eq4}
\end{equation}
Consequently, we get $\chi_c=n_h/ T$. Such a behavior is evident in
Fig.~4, quite universally for all $U>0$ and its validity extends at
lowest $T$ up to $n_h \sim 0.1$. A large increase in the maximum
$\chi_s$ at low $T$, being again rather insensitive to $U/t$, is a
clear manifestation of strong correlations and of the increasing
effective density of states ${\cal N}_F^c$ on approaching the
metal-insulator transition.  In fact, it has been claimed on the basis
of the $T=0$ QMC results \cite{imad} that within the Hubbard model the
charge susceptibility diverges approaching the half-filling as $\chi_c
\propto (1-n)^{-1/2}$. The latter is qualitatively consistent with the
flattening of the chemical potential as a function of doping $\mu(n
\to 1)$ in La$_{2-x}$Sr$_x$CuO$_4$ observed via the ARPES measurements
\cite{ino}. Nevertheless, at given lowest $T\sim T_{fs}\sim 0.1 t$ we
cannot distinguish a scenario with an enhanced but finite ${\cal
N}_F^c$ at $T=0$ from a divergent behavior.

Let us summarize some essential conclusions of the present study of
thermodynamic properties of the planar Hubbard model:

\noindent a) The FTLM seems to have advantages with respect to QMC and other
numerical methods for the calculation of thermodynamic quantities away
from half filling. The phase averaging method used in this study
represents an essential improvement and to large extent reduces
finite-size effects, in particular at moderate $U<W$.  Using the FTLM
and a phase averaging we reach in our study the low-$T$ regime,
i.e. $T<\tilde J$, where $\tilde J$ is an effective scale where the
spin exchange is fully active.

\noindent b) At large $U/t \gsim 12$ results for the thermodynamic
quantities match even quantitatively those of the corresponding
$t$-$J$ model (with $J\sim 4t^2/U$) \cite{jpterm,jprev} in the low
temperature $T<t$ window. Excitations into the upper Hubbard band
contribute significantly only at large $T>t$. On the other hand, for
smaller $U\lsim 8~t$ both scales start to merge, and become
inseparable for $n \neq 1$. In this intermediate $U$ regime there is
still a qualitative but not a quantitative resemblance to low-$T$
results within the $t$-$J$ model. Finally, results for $U \lsim 4~t$
approach the behavior of noninteracting fermions.

\noindent c) The effective exchange scale $\tilde J$ seems to
determine the 'optimal' doping for the entropy maximum
$s(n=1-n_h^*)=~$max as well as the pseudogap scale $T^*(n_h)$ in
$\chi_s(T^*)=~$max. It is clear that only at large $U$ we observe
$\tilde J \sim J \sim4 t^2/U$. On the other hand, for $U<12~t$ we see
that the positions of extrema in $s(n)$ and in $\chi_s(T)$ are quite
insensitive to $U/t$, indicating a rather constant $\tilde J \lsim
0.3~t $ as well as the 'optimum' doping $n_h^* \sim 0.15$. The former
fact can be understood in terms of less localized character of spin
degrees, which leads to an effective spin exchange interaction reduced
relative to the large $U$ expression $J=4t^2/U$.

\noindent d) The pseudogap feature (maximum) in $\chi_s(T)$ is well
visible at $U=12~t$, but remains only weakly pronounced at $U/t=8$ and
finally vanishes for smaller $U$. This is consistent with the
interpretation that the (large) pseudogap $T^*$ is related to an onset
of short-range antiferromagnetic correlations, which are only weakly
pronounced for $U\lsim 8~t$ away from half-filling $n < 1$.

\noindent e) One expects also an analogous pseudogap in the
specific-heat coefficient $\gamma(T)=C_V(T)/T$ \cite{coop}, where a
depletion should appear at $T<\tilde T^*(n)$. It is evident that such
an effect is present within the model, since near half-filling we see
$C_v \propto T^2$ while at larger doping $n_h \to n_h^*$ we get $C_v
\propto T^\nu$ with $\nu \leq 1$.
 
\noindent f) We should note that the maximum in $\chi_s(T)$ is not
specific for the 2D Hubbard model, but seems to be generally present
also in the 1D model \cite{degu}. Nevertheless, in a 1D system there
is no qualitative change in the character of low-energy (spin and
charge) excitations on doping since the excitations have all the way a
linear dispersion and consequently a nonvanishing $\gamma(T\to 0)$, in
contrast to a 2D system.

\noindent g) Previous studies of thermodynamic quantities within the
$t$-$J$ model \cite{jprev} have shown that results (at $J/t=0.3$) are
even quantitatively in agreement with the experimental ones in
hole-doped cuprates, in particular the doping dependence of the
entropy $s$ \cite{coop}, the spin susceptibility $\chi_s$ \cite{torr}
and chemical potential $\mu$ \cite{ino}.  Our results show essentially
equivalence of the low-$T$ behavior of the $t$-$J$ model and Hubbard
model with large $U\gg t$, hence the correspondence with experiments
applies again. However, we have shown that many results do not change
significantly in a broader range of $U$, i.e.  there is even a
quantitative similarity of $s(t)$, $n_h^*$, $T^*(n_h)$ etc., so the
agreement with experiments persists also in a broader range of $U/t$.

Authors acknowledge the support of the Ministry of Education, Science
and Sport of Slovenia.

\end{document}